# SCALAR FIELD MODELS OF DARK ENERGY WITH BAROTROPIC EQUATION OF STATE: PROPERTIES AND OBSERVATIONAL CONSTRAINTS FROM DIFFERENT DATASETS

B. Novosyadlyj, O. Sergijenko

Astronomical Observatory of Ivan Franko National University of Lviv
8, Kyryla i Methodia str., Lviv, 79005, Ukraine
E-mail: novos@astro.franko.lviv.ua, olka@astro.franko.lviv.ua

*The possibility of constraining the parameters of scalar field dark energy with barotropic equation of state using different available datasets is discussed. It has been found that the initial value of dark energy equation of state parameter is constrained very weakly by most of the data. We have determined the constraints on this parameter, which come from the combined dataset including supernovae from the full SDSS compilation with the MLCS2k2 fitting of light curves. We discuss also the possibility of distinguishing between different dark energy models with barotropic equation of state using the future data on CMB anisotropies.*

**Introduction**

The unknown nature of dark energy – the mysterious component accelerating the expansion of Universe – is one of the main puzzles of modern cosmology. Since the simplest explanation – the cosmological constant in Einstein equations – faces a lot of interpretational problems, a variety of alternative models have been proposed. One of the most popular alternative approaches treats the dark energy as a scalar field with a given Lagrangian. The model of such field is defined by its potential which can be either physically motivated or obtained via reverse engineering from the variables describing dark energy in phenomenological fluid approach: its energy density $\rho_{de}$ and equation of state parameter $w$. The latter one can either be constant or vary in time. The character of temporal variation of $w$ is usually assumed *ad hoc*. Nevertheless, the physically motivated dependences of the equation of state on time are sought.

The only way to verify the plausibility of a dark energy model is to confront its predictions with the observational data and to find the allowable ranges of its parameters. Today the most precise data are obtained by the WMAP satellite power spectra of CMB temperature fluctuations, but operating and future experiments (e.g. Planck satellite) will provide us with the sufficiently precise polarization data. Moreover, it should be worth to extract the weak lensing information from the CMB data with such precision. The role of weak lensing is to remap the observation direction from **n** to **n'=n+d(n)**, where **d(n)** is the weak lensing deflection angle.

The goal of this paper is to review the general properties of scalar field models of dark energy with barotropic equation of state, to discuss the possibility of constraining the parameters of such models using different current and expected data and to present the observational constraints on cosmological models with classical scalar field with barotropic equation of state as dark energy.

**Properties of scalar field models of dark energy with barotropic equation of state**

We assume that the background Universe is spatially flat, homogeneous and isotropic with Friedmann-Robertson-Walker (FRW) metric of 4-space $ds^2 = g_{ij} dx^i dx^j = a^2(\eta)(d\eta^2 - \delta_{\alpha\beta} dx^\alpha dx^\beta)$, where $\eta$ is the conformal time defined as $dt = a(\eta) d\eta$ and $a(\eta)$ is the scale factor, normalized to 1 at the current epoch $\eta_0$ (here and below we put $c=1$). The Latin indices $i, j,...$ run from 0 to 3 and the Greek ones are used for the spatial part of the metric: $\alpha, \beta, ... = 1, 2, 3$.

We consider the Universe filled with non-relativistic particles (cold dark matter and baryons), relativistic ones (thermal electromagnetic radiation and massless neutrino) and minimally coupled dark energy. The last one is assumed to be the scalar field with either Klein-Gordon (classical, below: CSF) or Dirac-Born-Infeld (tachyonic, below: TSF) Lagrangian

$$L_{clas} = X - U(\varphi), \qquad L_{tach} = -\tilde{U}(\xi)\sqrt{1 - 2\tilde{X}},$$

where $U(\varphi)$ and $\tilde{U}(\xi)$ are the field potentials defining the model of the scalar field, $X = \varphi_{;i}\varphi^{;i}/2$ and $\tilde{X} = \xi_{;i}\xi^{;i}/2$ are kinetic terms. We assume the homogeneity of background scalar fields ($\varphi(x,\eta) = \varphi(\eta)$, $\xi(x,\eta) = \xi(\eta)$), so their energy density and pressure depend only on time:

$$\rho_{clas} = X + U(\varphi), \qquad P_{clas} = X - U(\varphi),$$
$$\rho_{tach} = \tilde{U}(\xi)/\sqrt{1 - 2\tilde{X}}, \qquad P_{tach} = -\tilde{U}(\xi)\sqrt{1 - 2\tilde{X}}.$$

The equation of state (EoS) parameters $w \equiv P_{de}/\rho_{de}$ for these fields are following:

$$w_{clas} = \frac{X-U}{X+U}, \qquad w_{tach} = 2\tilde{X} - 1.$$

The dynamics of expansion of the Universe is completely described by the Einstein equations

$$R_{ij} - \frac{1}{2} g_{ij} R = 8\pi G \left( T_{ij}^{(m)} + T_{ij}^{(r)} + T_{ij}^{(de)} \right), \qquad (1)$$

where $R_{ij}$ is the Ricci tensor and $T_{ij}^{(m)}$, $T_{ij}^{(r)}$, $T_{ij}^{(de)}$ are the energy-momentum tensors of non-relativistic matter *(m)*, relativistic matter *(r)*, and dark energy *(de)* correspondingly. Assuming that the interaction between these components is only gravitational, each of them should satisfy the differential energy-momentum conservation law separately, which for the perfect fluid with density $\rho_n$ and pressure $P_n$, related by the equation of state $P_n = w_n \rho_n$, gives:

$$a\rho_n' = -3\rho_n (1 + w_n), \qquad (2)$$

here and below a prime denotes the derivative with respect to the scale factor *a*. For the non-relativistic matter $w_m = 0$ and $\rho_m = \rho_m^{(0)} a^{-3}$, for the relativistic one $w_r = 1/3$ and $\rho_r = \rho_r^{(0)} a^{-4}$. Hereafter "0" denotes the current values.

The EoS parameter *w* and the adiabatic sound speed $c_a^2 \equiv \dot{P}_{de} / \dot{\rho}_{de}$ of dark energy are related by the ordinary differential equation:

$$a w' = 3(1+w)(w - c_a^2). \qquad (3)$$

In the first case the repulsive properties of scalar fields are raising, in the second one – receding. In the general case $c_a^2$ can be an arbitrary function of time, but here we assume that it is constant: $c_a^2 = const$. In such case the temporal derivative of $P_{de}(\eta)$ is proportional to the temporal derivative of $\rho_{de}(\eta)$, or in integral form:

$$P_{de} = c_a^2 \rho_{de} + C, \qquad (4)$$

where *C* is a constant. The above expression is the generalized linear barotropic equation of state.

The solution of the differential equation (3) for $c_a^2 = const$ is following:

$$w(a) = \frac{(1+c_a^2)(1+w_0)}{1 + w_0 - (w_0 - c_a^2) a^{3(1+c_a^2)}} - 1, \qquad (5)$$

where the integration constant $w_0$ is chosen as the current value of *w*. One can easily find that (5) gives (4) with $C = \rho_{de}^{(0)}(w_0 - c_a^2)$, where $\rho_{de}^{(0)}$ is current density of dark energy. Substituting (5) into (3) we see that for quintessential fields ($w_0 > -1$) the derivative of EoS parameter with respect to the scale factor is negative for $w_0 < c_a^2$ and positive for $w_0 > c_a^2$.

Thus, we have two values $w_0$ and $c_a^2$ defining the EoS parameter *w* at any scale factor *a*. As it follows from (5), $c_a^2$ corresponds to the EoS parameter at the beginning of expansion ($w(0) \equiv c_a^2$, which we denote by $w_i$; in all expressions below instead of $c_a^2$ we use $w_i$ as it has more clear meaning). The differential equation (2) with *w* from (4) has the analytic solution too:

$$\rho_{de} = \rho_{de}^{(0)} \frac{(1+w_0) a^{-3(1+w_i)} + w_i - w_0}{1 + w_i}.$$

Using the obtained dependences of densities of each component on the scale factor the following equations for background dynamics can be deduced from the Einstein equations (1):

$$H = H_0 \sqrt{\Omega_r / a^4 + \Omega_m / a^3 + \Omega_{de} f(a)}, \qquad (6)$$

$$q = \frac{1}{2} \frac{2\Omega_r / a^4 + \Omega_m / a^3 + (1 + 3w) \Omega_{de} f(a)}{\Omega_r / a^4 + \Omega_m / a^3 + \Omega_{de} f(a)}, \qquad (7)$$

where $f(a) = \left[ (1+w_0) a^{-3(1+w_i)} + w_i - w_0 \right] / (1 + w_i)$. Here $H \equiv \dot{a}/a^2$ is the Hubble parameter

(expansion rate) for any moment of time and $q \equiv -\left(a\ddot{a}/\dot{a}^2 - 1\right)$ is the acceleration parameter. They completely describe the dynamics of expansion of the homogeneous isotropic Universe.

There are 3 possible variants of future evolution of the Universe defined by the relationship between $w_0$ and $w_i$ for given values of rest of the parameters $\Omega_m$, $\Omega_r$ and $\Omega_{de}$ [1]:

1. $w' < 0$ $(w_i > w_0)$: In this case $w$ decreases monotonically from $w_i$ at the initial stage of expansion to $w_0$ today up to -1 at the infinite time. The constant $C$ in EoS (4) is negative. The dark energy density and pressure tend asymptotically to $\rho_{de}^{(\infty)} = \rho_{de}^{(0)}(w_i - w_0)/(1 + c_a^2)$ and $P_{de}^{(\infty)} = -\rho_{de}^{(\infty)}$. Hence, in this case the scalar field rolls slowly to the vacuum and in far future the Universe will proceed into de Sitter stage of its expansion with $q^{(\infty)} = -1$ and $H^{(\infty)} = \sqrt{\Omega_{de}(w_i - w_0)/(w_i + 1)}\, H_0$.

2. $w' = 0$ $(w_i = w_0)$: It corresponds to the well-studied case $w = const$. Here $C=0$ which gives the usual barotropic EoS $P_{de} = w_0 \rho_{de}$, $\rho_{de} \to 0$ when $a \to \infty$. So, the future Universe will experience the power law expansion with $a \propto t^{2/3(1+w_0)}$ and acceleration parameter $q \to (1 + 3w_0)/2$.

3. $w' > 0$ $(w_i < w_0)$: In such case the EoS parameter $w$ increases monotonically from $w_i$ at the initial stage of expansion of the Universe to $w_0$ at the current one and still continues to increase after that. In the flat Universe it will reach 0 at $a_{(w=0)} = \left[w_i(1+w_0)/(w_i - w_0)\right]^{1/3(1+w_i)}$ and 1 at $a_{(w=1)} = \left[(1-w_i)(1+w_0)/(2(w_0-w_i))\right]^{1/3(1+w_i)}$. The densities of scalar fields at these $a$ are positive: $\rho_{de}(a_{(w=0)}) = \rho_{de}^{(0)}(w_i - w_0)/w_i$ and $\rho_{de}(a_{(w=1)}) = \rho_{de}^{(0)}(w_i - w_0)/(w_i - 1)$ correspondingly. The dark energy will satisfy the strong energy condition $\rho_{de} + 3P_{de} \geq 0$ starting from $a_{(q=0)} = \left[(1+w_0)(1+3w_i)/(2(w_i - w_0))\right]^{1/3(1+w_i)}$ and then the accelerated expansion of the Universe will be changed by the decelerated one. The density of scalar field continues decreasing, reaches 0 at $a_{(\rho=0)} = \left[(1+w_0)/(w_0 - w_i)\right]^{1/3(1+w_i)}$ and then becomes negative. The EoS parameter at this moment has the discontinuity of the second kind. Later, when $\rho_m + \rho_{de}$ reaches 0, the expansion of the Universe is changed by the contraction since at this moment $\dot{a} = 0$, $\ddot{a} < 0$, as it follows from equations (6) and (7) having no solution for larger $a$.

**Observational constraints on scalar field models of dark energy with barotropic equation of state**

The described scalar field model of dark energy involves 3 parameters $\Omega_{de}$, $w_i$ and $w_0$ which should be determined comparing the calculated predictions on dynamics of expansion and large scale structure of the Universe with corresponding observational data. Since all predictions and data are related with other components, the determination should be done jointly with other cosmological parameter.

We consider the cosmological model with minimal set of 6 parameters: baryons density parameter $\Omega_b$, cold dark matter density parameter $\Omega_{cdm}$, Hubble constant $H_0$, spectral index of initial matter density power spectrum $n_s$, amplitude of initial matter density power spectrum $A_s$ and reionization optical depth $\tau_{rei}$. So, we have 9 unknown parameters, but the number of independent ones is 8, since we have assumed the spatial flatness of the Universe. Indeed, the dark energy density parameter $\Omega_{de}$ in this case is obtained from the zero curvature condition: $\Omega_{de} = 1 - \Omega_b - \Omega_{cdm}$.

To determine the best fitting values and confidential ranges of the scalar field parameters together with other cosmological ones in our previous work [1] we have performed the Markov Chain Monte Carlo (MCMC) analysis for the set of current observational data, which include the power spectra from WMAP7 [2,3,4] and SDSS DR7 [5], the Hubble constant measurements [6], the light curves of SN Ia from Union2 compilation [7] and Big Bang nucleosynthesis (BBN) prior [8,9].

We use the publicly available package CosmoMC [10,11], which includes the code CAMB [12,13] for calculation of model predictions for sampled sets of 8 cosmological parameters listed above. The CosmoMC has been modified to be run with the proposed here parametrization of dark energy EoS parameter (5) and Hubble parameter (6). The flat priors $-1 < w_0, w_i \leq 0$ have been used to take into account the quintessential properties of scalar fields with classical and tachyonic Lagrangians (lower limit) and the constraints following from observational data related to the recombination and nucleosynthesis epochs (upper one).

We have performed two MCMC runs for the flat cosmological model with CSF. Each run had 8 chains and the number of samples in each chain was ~200000. For the first run only the mentioned above flat prior for $w_i$ has been used. The set of best fitting parameters obtained by this run is marked by $p_1$ and presented in the Table together with $1\sigma$ limits from the extremal values of the N-dimensional distribution. All parameters except $w_i$ are well constrained, the one-dimensional posterior and mean likelihood distributions are close and similar to Gaussian (half-Gaussian for $w_0$), $1\sigma$ ranges are narrow. The initial value of EoS parameter $w_i$ is essentially unconstrained: its $1\sigma$ range is wide and coincides practically with the prior range [-1, 0]. The mean likelihood and posterior are different and the likelihood is bimodal. The first peak is close to -1, another one to 0. The best fitting value of $w_i$ in the set $p_1$ corresponds to the first peak. In this case $w_i < w_0$ which means that the best fitting scalar field model of dark energy has receding repulsion properties (*w'>0*).

Table: The best fitting values of cosmological parameters and $1\sigma$ limits from the extremal values of the N-dimensional distribution in the model with CSF determined by the Markov Chain Monte Carlo technique using the available observational data. First column – model of dark energy with increasing EoS parameter, the combined dataset WMAP7 + BBN + HST + SDSS LRG DR7 + SN Union2 (for details see [1]), second one – model of dark energy with decreasing EoS parameter, the combined dataset WMAP7 + BBN + HST + SDSS LRG DR7 + SN Union2, third one – the combined dataset WMAP7 + BBN + HST + SN SDSS (MLCS2k2). The current Hubble parameter $H_0$ is in units $km/(s \cdot Mpc)$, the age of the Universe $t_0$ is given in Giga years. Here $\omega_b = \Omega_b h^2$, $\omega_{cdm} = \Omega_{cdm} h^2$, where $h = \dfrac{H_0}{100\, km/(s \cdot Mpc)}$.

| Parameters | $p_1$ | $p_2$ | $p_3$ |
|---|---|---|---|
| $\Omega_{de}$ | $0.72^{+0.04}_{-0.06}$ | $0.71^{+0.04}_{-0.05}$ | $0.70^{+0.06}_{-0.07}$ |
| $w_0$ | $-0.93^{+0.13}_{-0.07}$ | $-0.99^{+0.16}_{-0.01}$ | $-0.81^{+0.20}_{-0.19}$ |
| $w_i$ | $-0.97^{+0.97}_{-0.03}$ | $-0.05^{+0.05}_{-0.94}$ | $-1.00^{+0.99}_{-0.00}$ |
| $10\omega_b$ | $0.225^{+0.017}_{-0.013}$ | $0.225^{+0.017}_{-0.012}$ | $0.227^{+0.017}_{-0.014}$ |
| $\omega_{cdm}$ | $0.111^{+0.013}_{-0.012}$ | $0.113^{+0.010}_{-0.013}$ | $0.109^{+0.017}_{-0.015}$ |
| $H_0$ | $69.2^{+4.2}_{-5.1}$ | $68.6^{+4.7}_{-4.4}$ | $66.0^{+5.5}_{-5.1}$ |
| $n_s$ | $0.97^{+0.05}_{-0.03}$ | $0.97^{+0.05}_{-0.04}$ | $0.97^{+0.04}_{-0.04}$ |
| $\log(10^{10} A_s)$ | $3.07^{+0.11}_{-0.08}$ | $3.09^{+0.10}_{-0.09}$ | $3.07^{+0.11}_{-0.08}$ |
| $\tau_{rei}$ | $0.084^{+0.049}_{-0.037}$ | $0.090^{+0.043}_{-0.039}$ | $0.085^{+0.048}_{-0.035}$ |
| $-\log L$ | 4027.35 | 4027.51 | 3859.271 |

To obtain the best fitting parameters corresponding to the second peak of the likelihood distribution we have performed analogical run with additional condition $w_i > w_0$. The set of best fitting parameters determined by this run is marked by $p_2$ and presented in the Table. Now, the best fitting value of $w_i$ corresponds to the second peak of mean likelihood distribution. In this case $w_i > w_0$ and the best fitting scalar field model of dark energy has raising repulsion properties (*w'<0*).

Note that the large variation of $w_i$ does not change essentially other parameters: each parameter from the set $p_2$ is in the $1\sigma$ range of the corresponding one from the set $p_1$ and vice versa. The $-\log L$'s (last row of the Table) for both sets are very close. Moreover, in [1,14] we have found that the dark energy model degeneracy is double: in type of the dynamics of scalar field (receding-raising repulsion properties) and in its Lagrangian (classical-tachyonic).

So, the used dataset does not allow us to constrain the values of $w_i$ In order to analyze the possibility of constraining this parameter we have performed the MCMC runs for the combined datasets including the WMAP7 data on CMB anisotropy, the Hubble constant measurements, Big Bang nucleosynthesis (BBN) prior and SN data with different light curve fittings.

Here we analyze following SN compilations:
- Union [15]: 307 SN with fitting SALT [16];
- Union2 [7]: 557 SN with fitting SALT2 [17];
- SDSS [18]: full sample (288 SN) with fitting SALT2;
- SDSS [18]: full sample (288 SN) with fitting MLCS2k2 [18,19].

Each of these runs has 8 chains converged to $R-1<0.01$. As we see in Fig. 1, for all datasets including SN data with SALT and SALT2 (Union, Union2 and SDSS) light curves fittings the initial value of the EoS parameter remains unconstrained (different shapes of 1D and 2D posteriors and mean likelihoods), while for the dataset including SN data from SDSS compilation with MLCS2k2 light curve fitting posteriors and mean likelihoods are similar and both one-dimensional distributions have the shape of half-Gaussian. This means that the last dataset allows us to put observational constraints on $w_i$. The obtained from these data best fitting parameters and their $1\sigma$ limits from the extremal values of the N-dimensional distribution are presented in the Table and marked $p_3$. The considerably smaller value of $-\log L$ for this parameter set reflects simply the fact that this parameters are obtained from the other dataset.

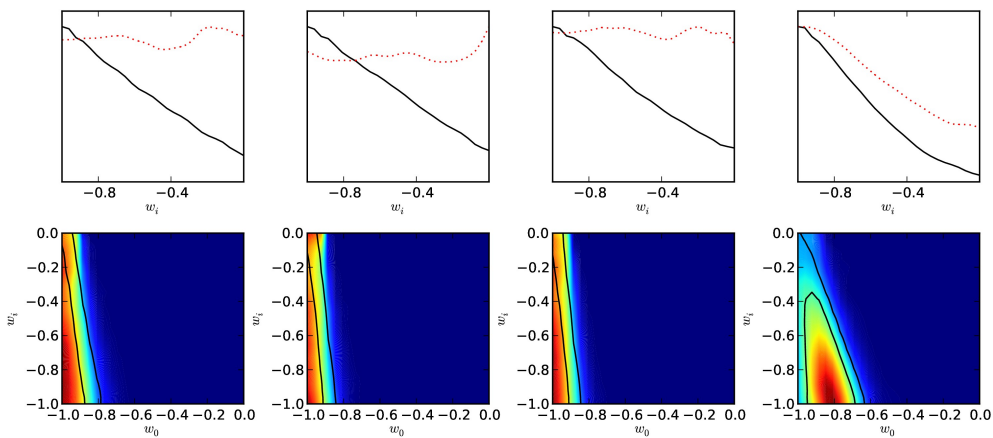

Fig. 1: Top: one-dimensional marginalized posteriors (solid lines) and mean likelihoods (dotted ones) for the combined datasets WMAP7 + BBN + HST + SN Union, WMAP7 + BBN + HST + SN Union2, WMAP7 + BBN + HST + SN SDSS (SALT2) and WMAP7 + BBN + HST + SN SDSS (MLCS2k2) (from left to right). Bottom: corresponding two-dimensional mean likelihood distributions in the plane $w_i - w_0$. Solid lines show the $1\sigma$ and $2\sigma$ confidence contours.

We have found that the data on SN Ia coming from the SDSS compilation with MLCS2k2 fitting of light curves allow the constraints on $w_i$ while the same data with SALT2 fitting do not. This is the manifestation of the well-known SALT2 vs MLCS2k2 discrepancy which is due mainly to the different rest-frame U-band models and assumptions about the source of color variations in both fitting methods [18]. We have performed the similar MCMC runs for the combined datasets including the SN subset NEARBY+SDSS (136 SN) from SDSS compilation, for which this discrepancy has been found to be the smallest [18]. As it can be seen in Fig. 2, in such case the parameter $w_i$ remains unconstrained for SN data with both light curve fitting. Therefore, the constraints on $w_i$ come from the treatment of higher-redshift SN samples by MLCS2k2 method, which differs from the corresponding treatment of SALT2 method.

The discussion of differences, benefits and limitations of SALT2 and MLCS2k2 light curve fitting methods is beyond the scope of this paper. However, since today it is not possible yet to determine definitely which fitting (if either) is better or incorrect, we regard our constraints on $w_i$ obtained from the combined dataset including SN data from the full SDSS compilation with MLCS2k2 fitting of light curves as the current observational constraints on this parameter. Thus the set of best fitting parameters and $1\sigma$ ranges $p_3$ represents the current observational constraints on spatially flat cosmological models with CSF with barotropic EoS as dark energy. It is worth noting that for the best fitting model $w_i < w_0$, so the field recedes its repulsion properties and in future the Universe will recollapse.

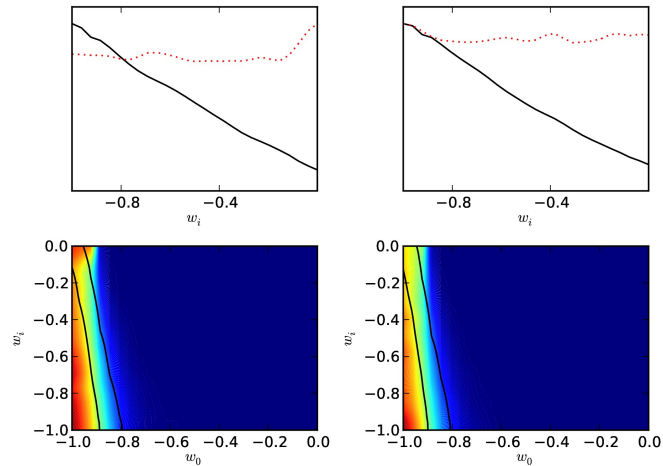

Fig. 2: Top: one-dimensional marginalized posteriors (solid lines) and mean likelihoods (dotted ones) for the combined datasets WMAP7+BBN+HST+SN SDSS (SALT2) and WMAP7+BBN+HST+SN SDSS (MLCS2k2) (from left to right) for the SN subset NEARBY+SDSS. Bottom: corresponding two-dimensional mean likelihood distributions in the plane $w_i - w_0$. Solid lines show the $1\sigma$ and $2\sigma$ confidence contours.

**Uncertainties of scalar fields with barotropic equation of state and expected Planck data**

We have seen that the currently available data do not allow the possibility to distinguish between the scalar field models of dark energy with different Lagrangians or characters of EoS parameter variation. To discuss such possibility for the future datasets let us compare the relative differences of CMB temperature fluctuations $\Delta C_l^{TT}/C_l^{TT}$, polarization $\Delta C_l^{EE}/C_l^{EE}$ and weak lensing deflection angle $\Delta C_l^{dd}/C_l^{dd}$ power spectra in models with both fields and parameter sets with the observational uncertainties, modelled for the Planck satellite in the following way. Assuming that the noise part is due to the combined effect of Gaussian beam and spatially uniform Gaussian white noise, for the experiment with known beam width and sensitivity the noise power spectrum for each channel can be approximated as follows:

$$N_l^{jj} = \theta_{fwhm}^2 \sigma_j^2 \exp\left[l(l+1)\frac{\theta_{fwhm}^2}{8\log 2}\right],$$

where $j$ stands for either *TT* or *EE*, $\theta_{fwhm}$ is the full width at half maximum of the Gaussian beam and $\sigma_j$ is the root mean square of the instrumental noise. The non-diagonal noise terms vanish since the noise contributions from different maps do not correlate. For the experiments with more than one channel the total noise power spectrum is obtained as:

$$\frac{1}{N_l^{jj(tot)}} = \sum_{i=1}^{n_{chan}} \frac{1}{N_l^{jj(i)}},$$

where $n_{chan}$ is the number of channels. The weak lensing deflection angle power spectrum is determined by the quadratic estimator method of Hu & Okamoto [20]. The described procedure was proposed by [21] and implemented in their code FuturCMB [22], which we use here.

In Fig. 3 we show the estimated errors for the Planck experiment with 3 channels (for each of them $\theta_{fwhm}$, $\sigma_T$ and $\sigma_E$ are 9.5 arcmin, 6.8 $\mu K$ per pixel and 10.9 $\mu K$ per pixel; 7.1 arcmin, 6.0 $\mu K$ per pixel and 11.4 $\mu K$ per pixel; 5.0 arcmin, 13.1 $\mu K$ per pixel and 26.7 $\mu K$ per pixel correspondingly). The observed sky fraction is assumed to be $f_{sky}=0.65$.

The inevitable uncertainties coming from the cosmic variance are calculated as follows:

$$\sigma_{(cv)_l} = \sqrt{\frac{2}{(2l+1)f_{sky}}} C_l.$$

We present the errors for the power spectra with points binned in the same progressively larger multipole bins with increasing $l$ as in [2] with means and errors computed as:

$$\langle C_l \rangle_i = \frac{\sum_{l'=l_{min\ i}}^{l_{max\ i}} C_{l'}}{N_i},$$

$$\frac{1}{\sigma_i^2} = \sum_{l'=l_{min\ i}}^{l_{max\ i}} \frac{1}{\sigma_{l'}^2}.$$

Here $l_{min\ i}$, $l_{max\ i}$, $N_i$ are the minimal, maximal $l$s and the total number of multipoles in the $i$th bin.

For each multipole the total uncertainty is determined in the following way:

$$\sigma_l^2 = \sigma_{(cv)l}^2 + N_l^2.$$

We see that at low spherical harmonics uncertainties of all spectra are cosmic variance dominated. At higher spherical harmonics the errors of polarization and lensing spectra are mainly instrumental noise dominated, while the ones of temperature fluctuations power spectrum remain cosmic variance dominated.

The models with the same fields but different parameter sets $p_1$ and $p_2$ can be distinguished by the data with such precision. For the CMB temperature fluctuations power spectrum the difference between studied models exceeds the estimated error level at high spherical harmonics, while for the polarization power spectrum at low spherical harmonics, where it is maximal. The models with different fields but the same parameter sets (corresponding to both decreasing and increasing EoS parameters) are still indistinguishable at such level of experimental precision [14]. The weak lensing data are completely irrelevant for the distinguishing between the studied models due to their limited precision as well as to the tiny differences of the corresponding theoretical spectra.

Note that if we modify the expression for $\sigma_l^2$ to match the used by WMAP team one [3]:

$$\sigma_l^2 = \sigma_{(cv)l}^2 + N_l^2 + 2C_l N_l,$$

we obtain higher error estimates. In such case the differences between power spectra in models with parameter sets $p_1$ and $p_2$ never exceed the error level, but they remain comparable with experimental uncertainties at high spherical harmonics for CMB temperature fluctuations (for TSF) and at low spherical harmonics for polarization (for both fields). Hence, the fields with increasing and decreasing EoS parameters would be distinguishable possibly by the Planck data alone and should be distinguishable by the combined datasets including them (see also the companion paper [14]).

**Conclusion**

We have studied the possibility of constraining the parameters of cosmological model with classical scalar field with barotropic equation of state as dark energy using the combined datasets including the power spectra from WMAP7, the Hubble constant measurements, Big Bang nucleosynthesis prior and the light curves of SN Ia from 3 different compilations: Union (SALT light curve fitting), Union2 (SALT2 light curve fitting) and SDSS (SALT2 and MLCS2k2 light curve fittings). We have found that the adiabatic sound speed, the parameter corresponding to the value of EoS at early epoch, is essentially unconstrained by the most of currently available data. For determination of the best fitting value and $1\sigma$ confidential ranges of $w_i$ the combined dataset including SN data from the full SDSS compilation with MLCS2k2 fitting of light curves should be used. The best fitting classical scalar field has the increasing EoS parameter and recedes its repulsion properties. We tried to determine the best fitting cosmological parameters and their $1\sigma$ confidential ranges using instead of the full SN SDSS compilation the subset NEARBY+SDSS for which the SALT2 vs MLCS2k2 discrepancy is the smallest and found that the possibility of constraining $w_i$ comes from the higher-redshift SN data with light curves fitted by MLCS2k2 method. We have also analyzed the possibility of distinguishing between classical and tachyonic scalar fields using the future data on CMB anisotropies from the Planck satellite. We found that the expected polarization data should be sufficient for distinguishing between the scalar fields with increasing and decreasing EoS parameters and the combined datasets including Planck data should allow us to put the stringent constraints on parameters of cosmological models with the scalar field with barotropic equation of state as dark energy.

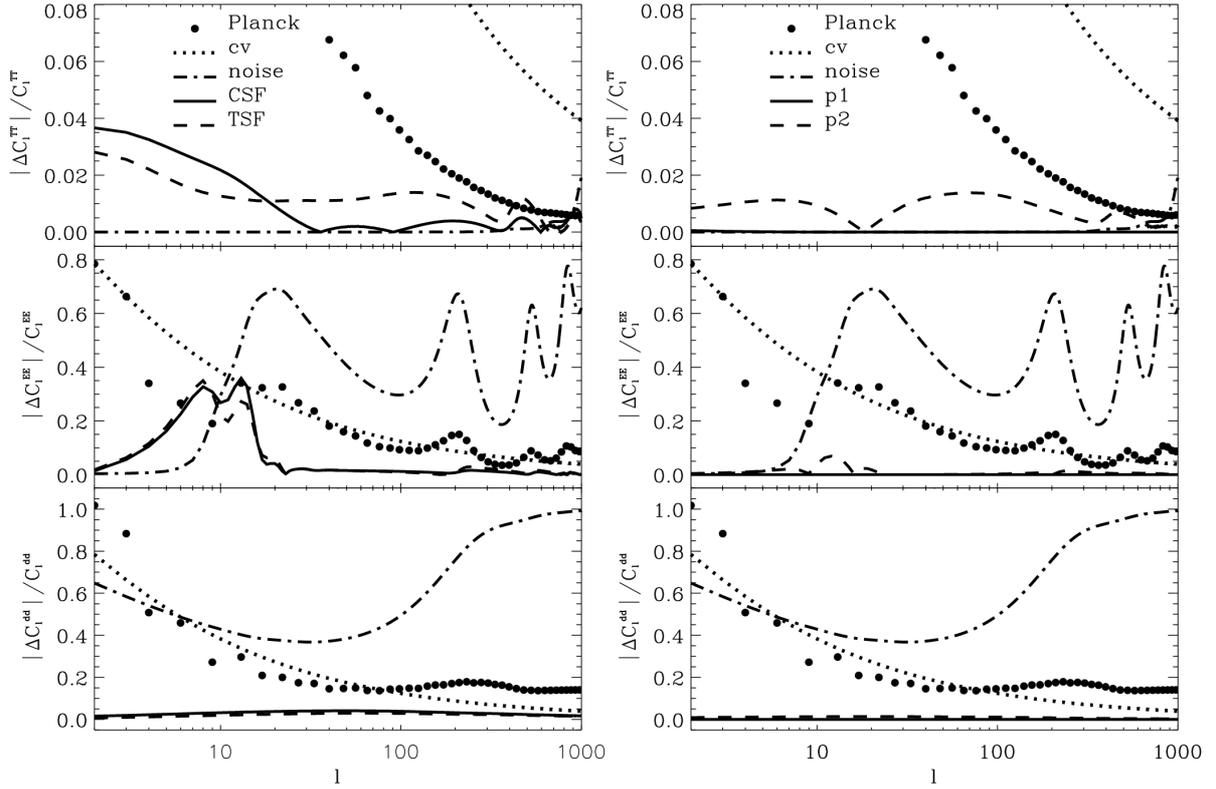

Fig. 3: Left: the relative differences of CMB temperature fluctuations $\Delta C_l^{TT}/C_l^{TT}$, polarization $\Delta C_l^{EE}/C_l^{EE}$ and weak lensing deflection angle $\Delta C_l^{dd}/C_l^{dd}$ power spectra (from top to bottom) in the models with classical and tachyonic scalar fields for two sets of the best fitting parameters $p_1$ and $p_2$. Right: the relative differences of CMB temperature fluctuations $\Delta C_l^{TT}/C_l^{TT}$, polarization $\Delta C_l^{EE}/C_l^{EE}$ and weak lensing deflection angle $\Delta C_l^{dd}/C_l^{dd}$ power spectra (from top to bottom) in the models with two sets of the best fitting parameters $p_1$ and $p_2$ for classical and tachyonic scalar fields. Dotted lines correspond to the cosmic variance, dash-dotted lines represent the noise-signal ratios, modelled for the Planck satellite. Circles show the estimated uncertainties of expected observational data.

This work was supported by the project of Ministry of Education and Science of Ukraine (state registration number 0110U001385), research program ``Cosmomicrophysics'' of the National Academy of Sciences of Ukraine (state registration number 0109U003207) and partially by the SCOPES project No. IZ73Z0128040 of Swiss National Science Foundation. Authors acknowledge the usage of CAMB, CosmoMC and FuturCMB packages. We are thankful to Main Astronomical Observatory of NASU and to High Performance Computing Centre of National Technical University of Ukraine 'Kyiv Polytechnic Institute' for the possibility to use the computer clusters for Markov Chain Monte Carlo runs.